\documentclass{article}

\usepackage{arxiv}

\usepackage[utf8]{inputenc} 
\usepackage[T1]{fontenc}    
\usepackage{hyperref}       
\usepackage{url}            
\usepackage{booktabs}       
\usepackage{amsfonts}       
\usepackage{nicefrac}       
\usepackage{microtype}      
\usepackage{graphicx}
\usepackage{tabularx}
\usepackage{gensymb}        
\usepackage{float}          
\graphicspath{ {./images/} }

\title{Assessment of design and analysis frameworks for on-farm experimentation through a simulation study of wheat yield in Japan}

\author{
 Takashi S. T. Tanaka \\
  Faculty of Applied Biological Sciences\\
  Gifu University\\
  1-1 Yanagido, Gifu, Japan 5011193 \\
  \texttt{takashit@gifu-u.ac.jp} \\
}

\begin{document}
\maketitle
\begin{abstract}
On-farm experiments can provide farmers with information on more efficient crop management in their own fields. Developments in precision agricultural technologies, such as yield monitoring and variable-rate application technology, allow farmers to implement on-farm experiments. Research frameworks including the experimental design and the statistical analysis method strongly influences the precision of the experiment. Conventional statistical approaches (e.g., ordinary least squares regression) may not be appropriate for on-farm experiments because they are not capable of accurately accounting for the underlying spatial variation in a particular response variable (e.g., yield data). The effects of experimental designs and statistical approaches on type I error rates and estimation accuracy were explored through a simulation study hypothetically conducted on experiments in three wheat fields in Japan. Isotropic and anisotropic spatial linear mixed models were established for comparison with ordinary least squares regression models. The repeated designs were not sufficient to reduce both the risk of a type I error and the estimation bias on their own. A combination of a repeated design and an anisotropic model is sometimes required to improve the precision of the experiments. Model selection should be performed to determine whether the anisotropic model is required for analysis of any specific field. The anisotropic model had larger standard errors than the other models, especially when the estimates had large biases. This finding highlights an advantage of anisotropic models since they enable experimenters to cautiously consider the reliability of the estimates when they have a large bias.
\end{abstract}

\keywords{Anisotropic variograms \and experimental design \and linear mixed models \and remote sensing \and sum-metric model}

\section{Introduction}
On-farm experimentation is a means of farmer-centric research and extension that examines the effect of crop management (e.g., fertilizer application, irrigation, and pest control) and variety selection on crop productivity in farmers’ own fields \cite{kindred2016agronomics, kyveryga2019farm}. Since the last century, agricultural experiments have been primarily performed by researchers in experimental fields under highly controlled conditions to ensure the accuracy of the estimated treatment effects. This typical research approach has contributed to an improved understanding of crop physiology and to developing agronomic practices, but the research results are not straightforwardly tailored to entire fields or regions. Thus, farmers and crop advisors have learned how to adjust crop management techniques by trial and error on farms \cite{sylvester1991modelling}. Developments in precision agricultural technologies, such as yield monitoring for combine harvesters and variable-rate application (VRA) technology (e.g., fertilizers, seeds and herbicides), allow farmers to run experiments on their own farms (Hicks et al. \cite{hicks1997analysis, pringle2004field1, marchant2019establishing}. On-farm experimentation has been gaining popularity by providing information on the best crop management techniques for specific regions, farmers, and fields \cite{kyveryga2019farm}. This increased popularity has identified new challenges for the implementation of on-farm experiments. \par \parindent = 20pt
To evaluate the precise effect of genotype and crop management on crop productivity, conventional small-plot experiments have been widely applied in agricultural research. These conventional small-plot experiments depend on the combination of three basic principles of experimental design (randomization, replication, and local control) and statistical approaches, such as analysis of variance (ANOVA), which were established by Fisher \cite{street1990fisher}. This fundamental agronomic research framework effectively separates the spatial variations and measurement errors from the observed data to detect the significance of the treatment effect \cite{kindred2016agronomics}. Widely used conventional statistical approaches, including ANOVA and ordinary least squares (OLS) regression, depend on the assumption that errors are independent. However, soil properties and crop yield are not spatially distributed at random, and similar values are observed near each other, which is called spatial autocorrelation \cite{mzuku2005spatial}. Spatial autocorrelation in a response variable (e.g., the crop yield) violates a conventional statistical assumption of independent errors, which leads to unreliable inferences (e.g., overestimation of the treatment effects) \cite{legendre2004effects,whelan2012small}. Thus, conventional statistical approaches are not directly applicable to on-farm experiments. In addition, these on-farm experiments are also characterized by generally having larger areas and simpler experimental arrangements. \par
To account for the underlying spatial structures, model-based geostatistics have been developed in disciplines associated with agricultural sciences \cite{oliver2010overview}; thus, geostatistical approaches have been applied to analyze on-farm data. For instance, yield data derived from chessboard or repeated strip trials have been kriged (interpolated) for other treatment plots, and a yield response model successfully established with a regression model \cite{pringle2004field2,kindred2015exploring}. However, it does not involve straightforward estimation with multiple kriging processes; it requires repeated and complicated experimental designs. While geostatistical approaches assumed stationarity of treatment effects and spatial autocorrelation, there is another approach based on no stationarity of treatment effects, namely spatially varying coefficient models \cite{trevisan2020spatial}. This approach can estimate treatment effects for each location, which can establish response-based prescriptions for VRA in large-scale fields. Not surprisingly, this approach cannot be implemented without VRA technology as it requires a completely randomized factorial design replicated over the entire field. Therefore, to be statistically robust, both approaches are reliant on complex experimental designs, which are invasive and complicated to design and implement. Consequently, farmers do not accept such approaches easily. \par
On-farm experimental designs must consider technical, agronomic and economic restrictions, but they also have to be statistically robust to provide reliable estimates of treatment effects \cite{whelan2012small}. A simulation study examined the effect of spatial structures, experimental designs and estimation methods on type I error rates and bias of treatment estimates by using geostatistical approaches \cite{alesso2019experimental}. This study indicated that higher spatial autocorrelation significantly increased type I error rates, while a spatial linear mixed model reduced them regardless of the experimental design, and more randomized and repeated experimental designs (e.g., split-planter, strip trials and chessboard designs) increased the accuracies of the treatment estimates. Furthermore, Marchant et al. \cite{marchant2019establishing} demonstrated that a spatial linear mixed model representing anisotropic spatial variations could successfully evaluate the treatment effect even in simple on-farm experimental designs (e.g. strip trials). Appropriate experimental designs and statistical approaches may vary according to the spatial variability and farmers’ available machinery, and a tradeoff between the simplicity of the experimental design and the desired precision of the outcome should be considered when conducting on-farm experiments \cite{alesso2019experimental}. \par
Previous on-farm experiments have been carried out in large-scale fields ranging from 8 to 16 ha in the UK \cite{marchant2019establishing} and from 10 to 100 ha in the US, South America and South Africa \cite{bullock2019data}. Given the recent worldwide commercialization of precision agricultural technologies, on-farm experiments can also benefit small-scale/smallholder farmers. However, appropriate and feasible experimental designs for small-scale fields may be different from those for large-scale fields due to the limitation of available areas per field or machine capability. Sensors measuring crop performance are the minimum requirement for the implementation of on-farm experiments. Low-cost commercial multispectral cameras mounted on unmanned aerial vehicles (UAVs) are currently increasingly available, not only for research use but also for crop advisory services. Crop yield can be predicted by UAV-based remote sensing \cite{zhou2020predicting,wang2014predicting}, even if the yield-monitoring combine harvesters are not affordable for small-scale/smallholder farmers. However, VRA technology, which is primarily used by large-scale farmers to implement on-farm experimental designs to achieve precise outcomes, remains in many places economically inaccessible. To be practical for small-scale/smallholder farms, farmers need to be able to perform on-farm experiments without VRA technology and the associated high investment cost. Therefore, the accumulation of more knowledge regarding the relationships between experimental designs and statistical approaches should be examined in small-scale fields. \par
The objectives of this study were to assess the effects of different sensor types, experimental designs and statistical approaches on type I error rates and estimation accuracy through a simulation study of on-farm experiments on small-scale wheat production in Japan. Furthermore, the inference framework for experimenters was examined from the perspective of model uncertainty. The predicted yield data were derived from remotely sensed imagery and commercial yield monitors for combine harvesters. Several hypothetical experimental designs were assumed to have been applied to those datasets. A spatial anisotropic model was developed to account for spatial autocorrelation and to reduce the estimation bias, and it was compared with OLS regression and standard spatial isotropic models as traditional statistical approaches.

\section{Materials and Methods}

\subsection{Yield data and experimental designs}
\begin{figure}[H] 
    \centering
    \includegraphics[width=0.65\textwidth]{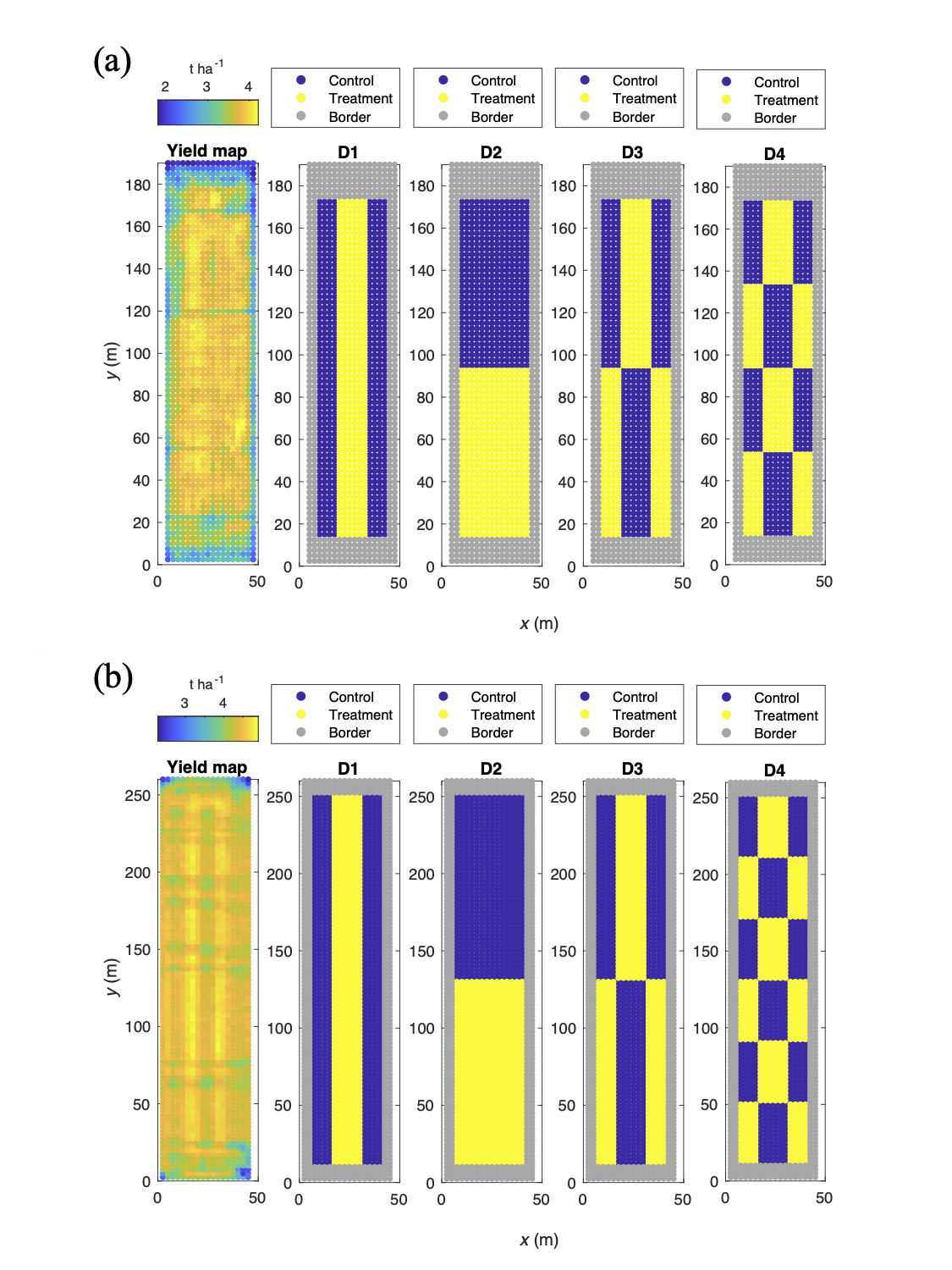}
    \caption{The yield map and simulated experimental designs for (a) Field 1 and (b) Field 2. Each yield point indicates an averaged value within 2.5-m grids. The border points (grey) indicate the area that was not used for the analysis due to potential edge effects.}
    \label{fig:fig1}
\end{figure}

Experimental designs and statistical approaches suitable for on-farm experiments in Japan were explored through a simulation study of winter wheat yield. Three fields were used for the simulation study. Currently, yield monitors on commercial combine harvesters are still not prevalent in Japan. Thus, crop yield maps of two upland fields converted from paddy fields (Fields 1 and 2) were derived from a previous study performed by Zhou et al. (2020) \cite{zhou2020predicting} (Figs.\ref{fig:fig1}). The fields were located in Sotohama, Kaizu, Gifu, Japan ($35{\degree}11$' N, $136{\degree}40$' E). The field sizes were $48\times180$ m (\textasciitilde0.86 ha) and $48\times260$ m (\textasciitilde1.25 ha) in Fields 1 and 2 respectively. These fields were remotely sensed by a commercial multispectral camera (Sequoia+, Parrot, France) mounted on a UAV at the grain-filling stage in 2018 and 2019. Briefly, winter wheat yield was predicted using a linear regression model with a predictor of enhanced vegetation index 2 (EVI2) \cite{jiang2008development} derived from the imagery. The ground sample distance of the imagery was 0.06 m pixel$^{-1}$. \par

\begin{figure}[H] 
    \centering
    \includegraphics[width=0.65\textwidth]{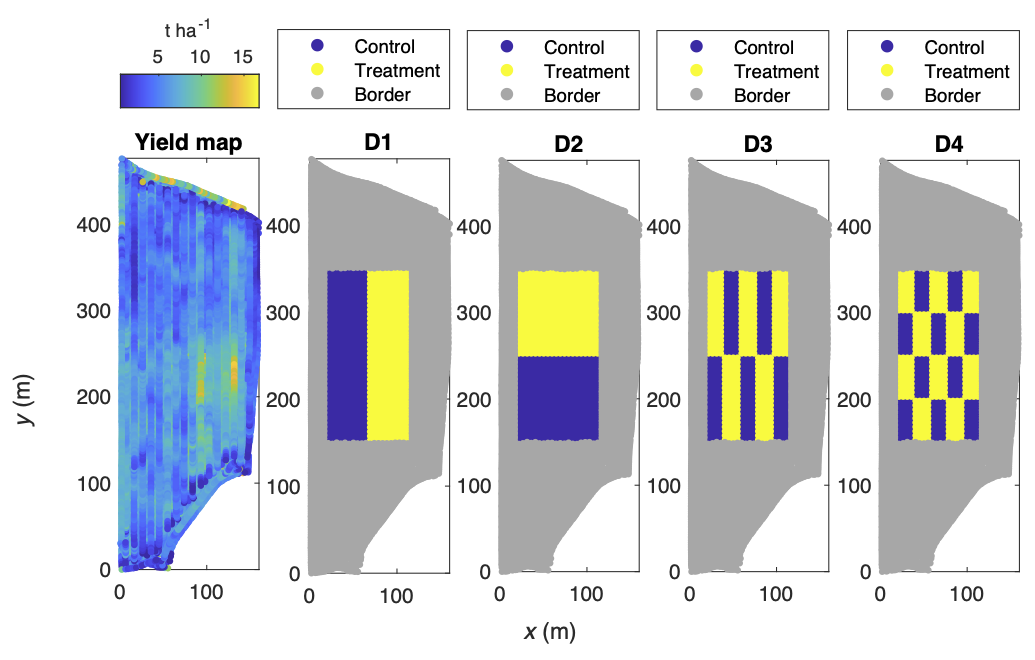}
    \caption{The yield map and simulated experimental designs for Field 3. The border points indicate the area that was not used for the analysis due to potential edge effects.}
    \label{fig:fig2}
\end{figure}

Yield monitor data were collected in 2016 at a demonstration farm (Field 3) of New Holland HFT Japan, Inc. with a commercial combine harvester CX8.70 (New Holland, Belgium) (Fig.\ref{fig:fig2}). The field was located in Tomakomai, Hokkaido, Japan ($42{\degree}45$' N, $141{\degree}44$' E). The entire field size was 8.45 ha, and the area used for the simulation was $100\times200$ m (2.00 ha). The header width was approximately 5 m, a typical size for a larger-scale farms in Hokkaido. The yield monitor on the combine harvester recorded the crop yield along harvest transects at an interval of approximately 1.3 m. \par

\subsection{Aggregation of yield data}
It is not feasible to use raw yield data in linear mixed models due to the high computational cost, particularly for anisotropic models implemented with a restricted maximum likelihood (REML) estimator (see statistical analysis section). Interpolated values at a certain grid size are generally used for mapping high-resolution spatial data. For Fields 1 and 2, the yield map derived from the UAV imagery was averaged within each square grid cell ($2.5\times2.5$ m) to generate mean yield values at the grid centroids. The raw yield monitor data in Field 3 was processed according to the method proposed by Marchant et al. (2019) \cite{marchant2019establishing} with modifications. Briefly, the co-ordinates were rotated as the combine harvester traveled in the $y$ direction. The $x$ co-ordinates were adjusted across each row in an exact straight line. Small variations in the co-ordinates and noise in the yield monitor data located near each other may prevent the estimation of the model parameters. In addition to the method proposed by Marchant et al. (2019) \cite{marchant2019establishing}, interpolation was further applied in this study since a square/rectangular lattice of points was preferred to fit the anisotropic model, as it could accommodate the computational cost and improve the precision of the maximum likelihood calculations. Therefore, the rotated yield monitor data points were averaged within each rectangular grid cell ($2.5\times5.0$ m in the $y$ and $x$ directions, respectively). For the interval of 2.5 m in the $y$ direction, each averaged yield data point contained 1--2 raw yield data points. 

\subsection{Simulation of effects on yield of different designs}
Two fertilizer treatments (control and treatment plots) were assumed to have been applied in the fields. Treatment plots received more fertilizer to theoretically increase yield. The experimental design might be very important to reduce the risk of type I errors and to evaluate the treatment effect precisely. From a practical viewpoint, a tractor equipped with an 18-m working width broadcaster was assumed to be used for the on-farm experiments, which is equivalent to 3 passes along the long-side direction for Fields 1 and 2 and is equivalent to 5 passes for Field 3. The specifications of this broadcaster were the same as local farmers'. The edges (15 m from edges associated with turning headlands and 6 m from edges parallel to field operations) were excluded from the analyses to avoid edge effects in Fields 1 and 2. Four experimental designs were simulated in Field 1 (Fig. \ref{fig:fig1} a), Field 2 (Fig. \ref{fig:fig1} b), and Field 3 (Fig. \ref{fig:fig2}). A simple strip trial (D1) was the easiest and most practical experimental design for farmers. A simple split-plot trial (D2) was established by splitting the experimental plots perpendicular to the farming operations, and it might require more complicated manual operation than D1. A combination of strip(s) and split-plot trials (D3) was established. A more repeated systematic design (D4) was established, which could not be implemented without VRA technology. Note that the analysis was performed through a simulation study, but the dataset was based on real collected data, which allows examination of the effect of anisotropy and unpredictable variations in the actual fields. 

\subsection{Statistical analysis using isotropic and anisotropic spatial linear mixed models}
OLS regression is based on the assumption that the errors are independent. In assessing the significance of the treatment effect on crop yield in on-farm experiments, spatial autocorrelation should be considered because the OLS estimator increases the risk of type I errors \cite{legendre2004effects, whelan2012small}. Thus, a spatial linear mixed model was used to evaluate the effects of hypothetical treatment on wheat yield. The spatial linear mixed model is written as

\begin{equation}
    y = X\beta + \epsilon
\end{equation}

where $y$ is a vector of length $n$ of the response variable, $n$ is the number of measurements, $X$ is the $n\times p$ fixed design matrix with the values of the vector of size $p$, which is the number of independent variables, $\beta$ is a vector of length $p$ of the fixed-effects coefficients, and $\epsilon$ is a vector of length $n$ of the random effects with covariance matrix $V$. The random effects are assumed to be spatially correlated, and the exponential function was used for the covariance estimation. The covariance function is written as 

\begin{equation}
    c(h) = \left\{ \begin{array}{ll}
    c_0 + c_1 & (h = 0) \\
    c_1 exp(-\frac{a}{h}) & (h>0)
    \end{array} \right.
\end{equation}

where $c_0$ is the nugget variance, $c_1$ is the sill variance, $h$ is the distance between the two measurements, and $a$ is the distance parameter. The theoretical variogram is written as

\begin{equation}
    r(h) = c_0 + c_1 (1 - exp(-\frac{a}{h}))
\end{equation}

The above model is a geometrically isotropic model, which has the same parameters in all directions. Marchant et al. (2019) \cite{marchant2019establishing} reported that yield monitor data showed strong anisotropy between the direction of the combine harvester's rows and perpendicular to the direction of the rows; thus, a product-sum covariance model \cite{de2001estimating} was used to model the variation along each direction. To establish an anisotropic model, direction-specific covariance functions should be parameterized. The sum-metric model presented by Bilonick (1988) \cite{bilonick1988monthly} was used in this study. The sum-metric model is written as

\begin{equation}
    c(h_x,h_y) = c_x(h_x) + c_y(h_y) + c_{xy}(h_{xy})
\end{equation}

\begin{equation}
    h_{xy} = \sqrt{h_x^2 + \alpha h_y^2}
\end{equation}

where $h_x$ is the lag perpendicular to the farming operation (across rows), $h_y$ is the lag in the direction of travel (within rows), and $h_{xy}$ is the lag obtained by introducing a geometric anisotropy ratio $\alpha$. The sum-metric model has been used for fitting space-time variograms previously \cite{snepvangers2003soil}, and its advantage is that the combination of 2-directional static components and 1 dynamic component of a covariance function is easily interpretable in a physical sense \cite{heuvelink1997spatio}. Thus, the isotropic model has three parameters to be estimated, and the anisotropic model has 8 parameters to be estimated. For the estimation of these random effects parameters, the REML estimator was used. The REML estimator does not depend on the unknown fixed effects; therefore, the estimates are less biased than maximum likelihood estimates \cite{cressie2015statistics}. Consequently, the REML estimator calculated the fixed-effect coefficient $\beta$ and its standard error. The statistical significance of the experimental treatment was assessed by $z$ statistics, and two-sided $p$-values were computed. The preferred model was evaluated based on the lower values of the Akaike information criterion (AIC) \cite{akaike1998information} between the geometrically isotropic and anisotropic models. Note that the REML estimator often has a risk of finding a non-optimum local solution as the optimization solver depends on initial values. Therefore, multiple initial values were used to iterate the REML estimation although it was computationally demanding. Most of the available applications for model-based geostatistics, such as the geoR package \cite{ribeiro2007geor} implemented in the R environment \cite{team2013r}, can only fit isotropic variograms based on the REML estimator. The R package gstat \cite{pebesma2004multivariable} is available for fitting space-time variograms, but the REML estimator is not implemented. Therefore, it was necessary to develop a user-friendly application that can fit an anisotropic model based on the REML estimator. All of the computations were conducted using MATLAB \cite{matlab}, and the documented MATLAB source code is available at GitHub (\url{https://github.com/takashit754/geostat}). Finally, the performances of the three models were compared: OLS regression models, spatial isotropic linear mixed models, and spatial anisotropic linear mixed models. For the spatial linear mixed models, the residual fitted variograms were evaluated to check the underlying spatial variations.

\subsection{Calculation of type I error rates and estimated accuracy}
To assess the estimated model accuracy, randomly generated numbers from a Gaussian distribution ($\mu$=0.3; $\sigma$ = 0.10 t ha$^{-1}$) were added to the yield data for each point in the treatment plots. Then, the bias was estimated by computing the difference between the fixed-effect coefficient $\beta$ from each statistical model and the population parameter yielded by a Gaussian random number generator (approximately 0.3 t ha$^{-1}$). In ideal circumstances, on-farm experiments with multiple treatments in a row achieved the standard error in winter wheat yield of less than 0.05 t ha$^{-1}$  \cite{marchant2019establishing}, which was equivalent to a least significant difference of < 0.1 t ha$^{-1}$. The least significant difference ranged from 0.4 to 0.8 t ha$^{-1}$ for small-plot field experiments on wheat crops in Japan \cite{nakano2008effect, okami2016analysis}. It is important to examine whether the small treatment effect can be detected precisely while avoiding the risk of type I error because such information is crucial to support the farmers' management decision. Therefore, the hypothetical treatment effect was intentionally set as the small value. In addition, 95\% confidence intervals were calculated using the standard error estimated from each model. The simulated type I error rates were presented as p-values of the experimental treatment on the assumption that the treatment population parameter was zero. To assess the effect of the experimental designs and statistical models on either the simulated type I error rates or the absolute bias, two-way ANOVA based on type III sums of squares was performed.

\section{Results}
\subsection{Simulated type I error rates}
The effects of the experimental design and model on the simulated type I error rates were evaluated by two-way ANOVA (Table~\ref{tab:table1}). There was a significant effect of model selection on the simulated type I error rates. There was no significant interaction between the experimental design and the model. The mean value of the simulated type I error rates was significantly higher in the anisotropic model than the in OLS model. The simulated type I error rates for each field are shown in Table 2. The simulated type I error rates were greater than the significance level (>0.05) in the simplest design (D1) for Field 1 (Table~\ref{tab:table2}). Moreover, they were greater than the significance level (>0.05) in the most repeated and complicated design (D4) in Fields 2 and 3 (Table~\ref{tab:table2}). Overall, lower simulated type I error rates were more frequently observed in the OLS model than in the other models. For the anisotropic model, there were no significant simulated type I error rates for Field 3. \par

\begin{table}[H]
 \caption{Mean values of the simulated type I error rates as affected by experimental design and model.}
  \centering
  \begin{tabular}{llll}
    \toprule
     & \multicolumn{2}{c}{Type I error rate} \\
    \midrule
    Design &  &  \\
    \multicolumn{1}{c}{D1} & 0.417 &  \\
    \multicolumn{1}{c}{D2} & 0.396 &  \\
    \multicolumn{1}{c}{D3} & 0.171 &  \\
    \multicolumn{1}{c}{D4} & 0.207 &  \\
    Model & & \\
    \multicolumn{1}{c}{OLS} & 0.153 & b \\
    \multicolumn{1}{c}{Isotropic} & 0.316 & ab \\
    \multicolumn{1}{c}{Anisotropic} & 0.425 & a \\
      ANOVA & & \\
    \multicolumn{1}{c}{Design} & n.s. & \\
    \multicolumn{1}{c}{Model} & * & \\
    \multicolumn{1}{c}{Design$\times$Model} & n.s. & \\
    \bottomrule
  \end{tabular}
  \\Different small letters indicate significant difference ($p$ < 0.05). *: $p$ < 0.05; n.s.: not significant.
  \label{tab:table1}
\end{table}

\begin{table}[H]
 \caption{The simulated type I error rates for Fields 1, 2, and 3.}
  \centering
  \begin{tabular}{p{1cm}p{1cm}p{1.5cm}p{1.8cm}p{1cm}p{1.5cm}p{1.8cm}p{1cm}p{1.5cm}p{1.8cm}}
    \toprule
    & \multicolumn{3}{c}{Field 1} & \multicolumn{3}{c}{Field 2} & \multicolumn{3}{c}{Field 3}   \\
    \cmidrule(r){2-4}
    \cmidrule(r){5-7}
    \cmidrule(r){8-10}
    Design & OLS model & Isotropic model & Anisotropic model & OLS model & Isotropic model & Anisotropic model & OLS model & Isotropic model & Anisotropic model \\
    \midrule
    D1 & 0.17 & 0.89 & 0.90 & \textbf{<0.01} & 0.60 & 0.54 & \textbf{<0.01} & \textbf{<0.01} & 0.65 \\
    D2 & 0.20 & 0.10 & 0.93 & \textbf{0.03} & 0.31 & 0.44 & \textbf{<0.01} & 0.90 & 0.67\\
    D3 & \textbf{<0.01} & 0.59 & \textbf{0.01} & 0.17 & 0.22 & \textbf{<0.01} & \textbf{<0.01} & \textbf{0.03} & 0.52 \\
    D4 & \textbf{0.03} & 0.10 & \textbf{0.01} & 0.89 & \textbf{<0.01} & 0.28 & 0.42 & 0.06 & 0.17 \\
    \bottomrule
  \end{tabular}
  \\The $p$-values in bold indicate statistical significance at the 0.05 significance level. 
  \label{tab:table2}
\end{table}

\subsection{The precision of estimates}
The effects of experimental design and model on the bias were evaluated by two-way ANOVA (Table~\ref{tab:table3}). The experimental design and model selection had no significant effects on the bias. However, there was a tendency that more complexity in experimental design and model decreased bias. \par

\begin{table}[H]
 \caption{Mean values of the estimated bias as affected by the experimental design and model.}
  \centering
  \begin{tabular}{llll}
    \toprule
     & \multicolumn{1}{c}{Bias} \\
    \midrule
    Design &  &  \\
    \multicolumn{1}{c}{D1} & 0.281 \\
    \multicolumn{1}{c}{D2} & 0.209 \\
    \multicolumn{1}{c}{D3} & 0.093 \\
    \multicolumn{1}{c}{D4} & 0.056 \\
    Model & & \\
    \multicolumn{1}{c}{OLS} & 0.233 \\
    \multicolumn{1}{c}{Isotropic} & 0.165 \\
    \multicolumn{1}{c}{Anisotropic} & 0.083 \\
      ANOVA & & \\
    \multicolumn{1}{c}{Design} & n.s. \\
    \multicolumn{1}{c}{Model} & n.s.  \\
    \multicolumn{1}{c}{Design$\times$Model} & n.s. \\
    \bottomrule
  \end{tabular}
  \\n.s.: not significant.
  \label{tab:table3}
\end{table}

\begin{figure}[H] 
    \centering
    \includegraphics[width=1.0\textwidth]{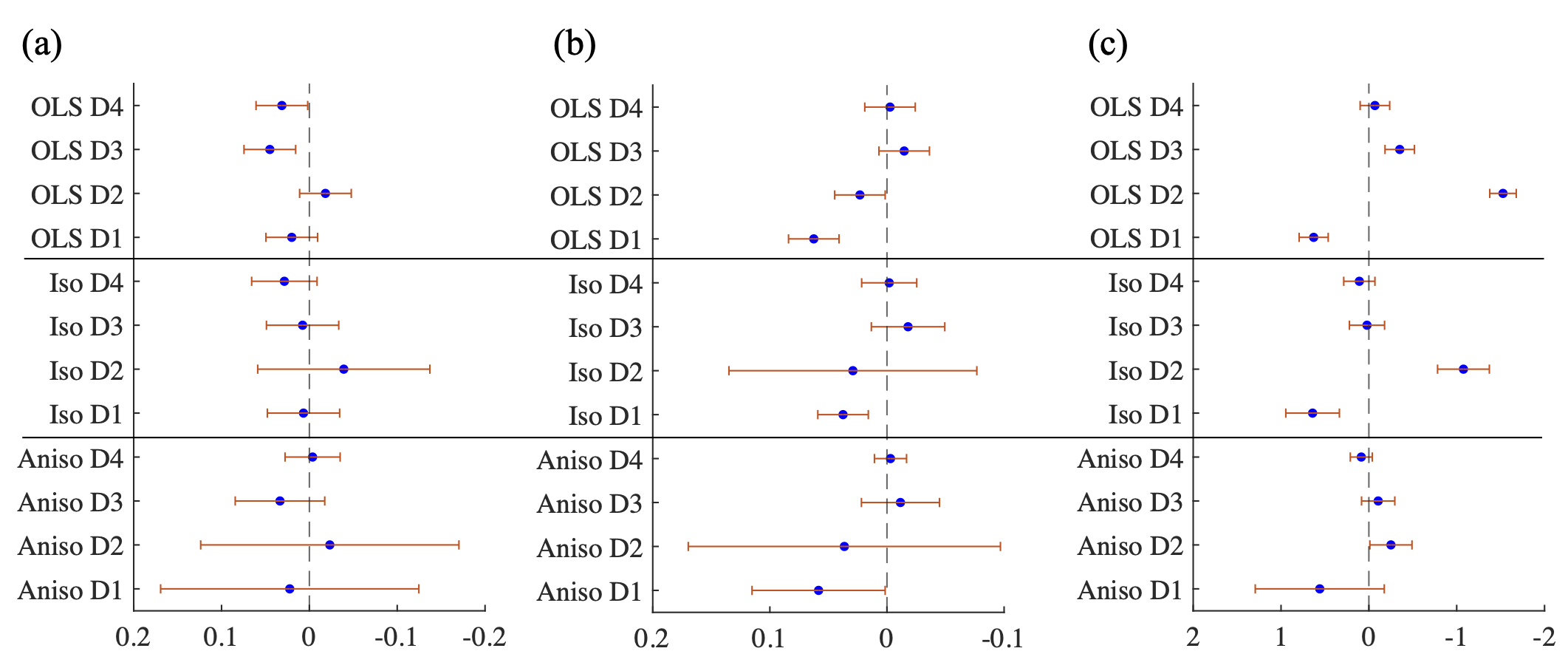}
    \caption{Bias and 95\% confidence intervals in Field 1 (a), Field 2 (b), and Field 3 (c). Iso and Aniso represent the isotropic and anisotropic models, respectively. }
    \label{fig:fig3}
\end{figure}

The bias and 95\% confidence intervals are shown in Fig. \ref{fig:fig3}. The closer the bias is to zero, the closer the estimates are to the actual treatment values yielded by the Gaussian random number generator (approximately 0.3 t ha$^{-1}$). Overall, the OLS model had narrower 95\% confidence intervals, and the 95\% confidence intervals did not contain zero more frequently than the other models. Furthermore, the 95\% confidence intervals of the OLS model did not contain zero, even in the most repeated and complicated design (D4) of Field 1. The anisotropic model had wider 95\% confidence intervals, especially for the simpler designs (D1 and D2) for all fields.

\subsection{Residual Variogram Analysis}
\begin{figure}[H] 
    \centering
    \includegraphics[width=1.0\textwidth]{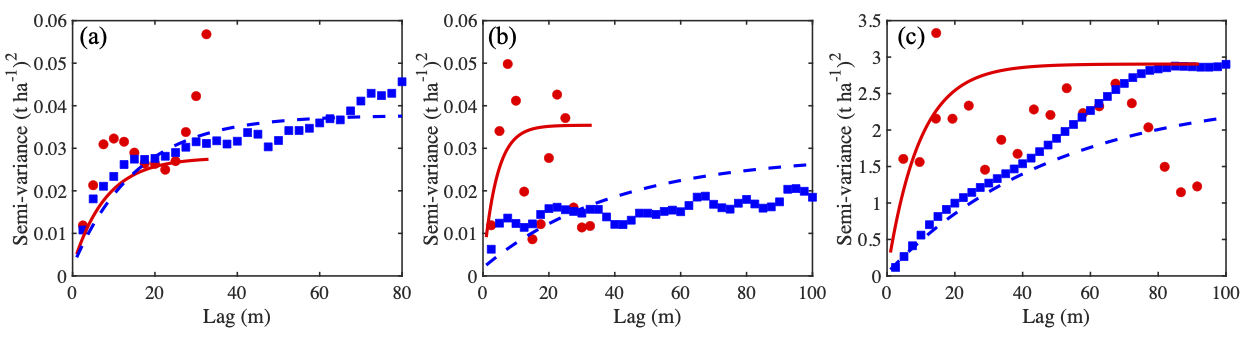}
    \caption{The anisotropic experimental and fitted residual variograms for Field 1 (a), Field 2 (b) and Field 3 (c). All the variograms were computed from the dataset for the estimation of simulated type I error rates in D1. Blue squares represent experimental variograms in the direction of farming (y). Red circles represent experimental variograms in the direction perpendicular to the farming operations (x). Experimental semi-variance with more than 30 pairs of identical lags are displayed. Blue dashed lines indicate fitted variograms in the direction of farming. Red solid lines indicate fitted variograms in the direction perpendicular to the farming operations.}
    \label{fig:fig4}
\end{figure}

The 2-directional residual variograms for the estimation of simulated type I error rates in D1 are shown in Fig. \ref{fig:fig4}. In Field 1, the sill variance in the fitted variogram was 0.036 (t ha$^{-1}$)$^2$ in the direction of farming operations ($y$), while it was 0.026 (t ha$^{-1}$)$^2$ in the direction perpendicular to the farming operations ($x$). The range parameter that reaches the sill variance at the 95\% level was two times larger in the direction of farming operations ($y$) (46.1 m) than in the direction perpendicular to the farming operations ($x$) (24.1 m). In contrast, in Field 2, the sill variance in the fitted variogram was approximately 0.034 (t ha$^{-1}$)$^2$ in the direction of farming operations ($y$), while it was 0.027 (t ha$^{-1}$)$^2$ in the direction perpendicular to the farming operations ($x$). The range parameter that reaches the sill variance at the 95\% level was 120 m in the direction of farming operations ($y$), while it was approximately 10 times larger than in the direction perpendicular to the farming operations ($x$) (12.3 m). In Field 3, the sill variance in the direction perpendicular to the farming operations ($x$) was 2.47 (t ha$^{-1}$)$^2$, and the sill variance in the direction of the farming operations ($y$) was 2.87 (t ha$^{-1}$)$^2$. The range parameter that reaches the sill variance at the 95\% level was 172 m in the direction of farming operations ($y$), while it was approximately 5 times larger than in the direction perpendicular to the farming operations ($x$) (31.7 m). Overall, there was strong anisotropy for Field 2 and 3.

\section{Discussion}
Sensor types, data preprocessing, experimental designs, statistical approaches, and within-field spatial structures affect the precision in on-farm experiments \cite{alesso2019experimental,marchant2019establishing}; thus, there are many possibilities for the best experimental design and statistical approaches in on-farm experiments \cite{kyveryga2019farm}. Those effects on the type I error rates and estimated accuracy were explored through a simulation study that generated hypothetical treatments on real wheat yield data in Japanese fields. Several important implications for experimenters can be drawn from the results. \par
The experimental designs did not significantly affect the simulated type I error rates although the more repeated and complicated design (D3 and D4) showed two times smaller simulated type I error rates (Table~\ref{tab:table1}). The results partially agreed with a previous study, which reported that designs with fewer replications and larger experimental units tended to increase the risk of type I error \cite{alesso2019experimental}. However, the simulated type I error rate was relatively higher in the simpler design (D1 and D2) than in the other more repeated designs (D3 and D4) in Field 1 (Table~\ref{tab:table2}). This contradiction may have occurred because the control plots of D3 and D4 coincided with low-yield areas in Field 1 (Fig. \ref{fig:fig1} a). These results indicated that repeated designs are not sufficient to avoid the risk of type I error. Although randomization in the repeated designs were not tested in this study, it may contribute to improving the precision of the experiments as it can accommodate both systematic and erratic spatial trends \cite{piepho2013why}. The effect of randomization on the precision of the experiment should be examined in further studies. However, it may not be practical to implement complex randomization with a variety of treatments in relatively small fields, particularly in Asian countries, even if VRA technology is available. Moreover, the model was a significant factor affecting the simulated type I error rates (Table~\ref{tab:table1}). The simulated type I error rates were significantly greater in the anisotropic model than in the OLS model. Similarly, the more repeated designs and anisotropic models tended to show smaller bias (Table~\ref{tab:table3}), particularly for Field 3 (Fig. \ref{fig:fig3}), although the effects were not significant. Consequently, the combination of a repeated design and an anisotropic model might sometimes be a solution for avoiding the risk of type I error and reducing the estimation bias. However, the isotropic model was able to avoid the risk of type I error for all designs in Field 1 (Table~\ref{tab:table2}). In this case, the AIC was less in the isotropic model than in the anisotropic model, but only for Field 1 (data not shown). These results showed that the best model can vary for different fields, according to the experimental design and field conditions. Therefore, model selection is important for obtaining robust outcomes through on-farm experiments. For instance, model selection should be carried out to determine whether the anisotropic model is required for analysis of any specific field. \par
The fitted residual variograms showed strong anisotropy with different sill variances between the directions for Field 2 and Field 3 rather than for Field 1 (Fig. \ref{fig:fig4}). The bias was greatly reduced by using an anisotropic model in D2 for Field 3. Furthermore, the anisotropic model had 95\% confidence intervals that contained zero for three designs (D1, D2, and D3), although OLS and isotropic models had 95\% confidence intervals that contained zero only for D4 and D2, respectively. These results are in agreement with the finding from Marchant et al. (2019) \cite{marchant2019establishing}, who demonstrated that isotropic models were sufficient for analyzing remotely sensed data but were not appropriate to account for spatial autocorrelation in yield monitor data. For yield data, they applied a product-sum model \cite{de2001estimating}, which is more complex than the standard isotropic model. This study demonstrated that the sum-metric model \cite{bilonick1988monthly} separated the underlying spatial variation not only from the yield monitor data (Field 3) but also from remotely sensed data (Field 2) to evaluate the treatment effects more accurately than the isotropic model. Thus, an anisotropic model is sometimes recommended for the analysis of on-farm data, even data derived from remotely sensed imagery. \par
In Field 2, experimental variograms indicated that semi-variance was not successfully explained only by lag distance in the direction perpendicular to the farming operations ($x$) (Fig. \ref{fig:fig4}). The yield data was noisy and independent according to the specific rows. This may be attributed not only to the direction of farming operations but also to the field size, shape or machine capability. Typical Japanese paddy fields are rectangular and small (e.g. less than 1 ha), so a narrow side may not be sufficiently large to spread fertilizer evenly. For instance, in Fields 1 and 2, the length on the narrow side was only 48 m, which cannot be divided by the broadcaster's working width (18 m). As two high-yield lines were indicated in Field 2 (approximately 15 and 30 m on the $x$-axis) (Fig. \ref{fig:fig1} b), some areas may have received fertilizer application twice, as suggested by Tanaka et al. (2019) \cite{tanaka2019assessing}. Therefore, small-scale paddy fields may inherently contain a high variability in the direction perpendicular to the farming operations if the automatic section control system in the broadcaster or sprayer is not used for on-farm experiments. It is noteworthy that these findings may be specific to small-scale fields in Asian countries. One solution might be data trimming before statistical analysis if such effects and errors could be identified from the experimenters' knowledge. Another solution might be to establish models with variable intercepts as a random effect for each row. However, the outcomes should be carefully interpreted as the model becomes complex. Further research is still required to confirm the factors underlying variability in yield and to explore the best research framework to implement on-farm experiments appropriate for Asian countries with relatively small fields. \par
To reiterate, an anisotropic model is an incomplete solution but provides more robust outcomes than traditional statistical approaches. An anisotropic model is advantageous since it covered larger standard errors when the estimates had large biases (Fig. \ref{fig:fig3}). Moreover, the 95\% confidence intervals of the OLS and isotropic models were generally narrow, and the hypothetical treatment effects fell outside of them. To make a reliable decision according to the results of on-farm experiments, experimenters should keep in mind that estimates may not always be precise, but they can consider how much it could vary, as this variability would result in an adverse scenario. Farmers are not interested in whether the treatment is significant at the 0.05 level but rather in whether there will be a return on investment \cite{kindred2016agronomics}. Therefore, as previous studies have determined \cite{kahabka2004spatial, whelan2012small}, it is necessary to examine economic feasibility (e.g., marginal profits) by using real on-farm data in further studies.

\section{Conclusions}
The outcomes of on-farm experiments can support farmers' decision-making processes, while inappropriate procedures will result in incorrect interpretations. The repeated experimental designs examined here did not contribute to reducing the risk of type I error and bias. Although there are many choices for the experimental design and statistical approaches, the combination of repeated designs and anisotropic models sometimes provides more reliable outcomes than the other methods to avoid issues arising from on-farm experiments. The results of the anisotropic model showed large standard errors, especially when the estimates had large biases. Considering that the aim of on-farm experiments is to provide farmers with information on economic feasibility, these statistical characteristics of anisotropic models are advantageous, as experimenters have opportunities to infer the analytical results conservatively. To examine the effect of these statistical characteristics on farmers' decision-making processes, economic analysis is needed using real on-farm data in the future. Overall, this study indicated that the basic framework of on-farm experiments such as experimental designs and statistical approaches, which has been originally developed in large-scale system, may also be applicable for small-scale system. However, this simulation study only examined three fields. Further research should be oriented towards exploring the factors underlying variability in yield and the best research framework to implement on-farm experiments appropriate for small-scale/smallholder farmers.

\section*{Acknowledgments}
The authors wish to thank the farming company 'Fukue-eino' for allowing the survey of their fields and New Holland HFT Japan, Inc. for providing the yield monitor data. This work was supported by the KAKENHI Early Career Scientists Grant from the Japan Society for the Promotion of Science (JSPS) (grant number 18K14452) and by the research grant from the Koshiyama Science and Technology Foundation.

\bibliographystyle{unsrt}  

\end{document}